# Effect of isothermal holding temperature on the precipitation hardening in Vanadium-microalloyed steels with varying carbon and nitrogen levels

A. Karmakar[1*], A. Mandal[1], S. Mukherjee[2], S. Kundu[2], D. Srivastava[3], R. Mitra[1], D. Chakrabarti[1]

[1]Department of Metallurgical and Materials Engineering, Indian Institute of Technology Kharagpur, Kharagpur, 721 302, India.

[2]Product Research Group, Research Development and Scientific Services, Tata Steel, Jamshedpur, 831 001, India.

[3]Materials Science Division, Bhabha Atomic Research Centre (BARC), Mumbai 400 085, India.

*Corresponding author email: *anish.met@gmail.com*

**Abstract:**

Combined effect of C and N levels and isothermal holding temperature on the microstructure, precipitation and the tensile properties of Vanadium (V) microalloyed steels (0.05 wt%. V) were studied. Two different V steels, one having higher C and lower N content, HCLN steel, and the other having lower C and higher N content, LCHN steel, were prepared and subjected to isothermal holding treatment over a temperature range of 500–750 °C, after hot-deformation. Maximum precipitation strengthening from fine V(C,N) precipitates has been found at intermediate isothermal holding temperatures (600–650 °C) in both the steels. In spite of the significantly smaller fraction of pearlite and bainite, coarser average ferrite grain size and lower precipitation-dislocation interaction in the ferrite matrix, the yield strength of LCHN steel was close to HCLN steel. This can be attributed to the higher precipitation strengthening (by ~ 20–50 MPa) resulted from the finer V-precipitates in LCHN steel than that of HCLN steel.

1.      **Introduction:**

Vanadium microalloyed high strength low alloy (HSLA) steels are widely being used for construction, line pipe, pressure vessel, automobile, naval and defense applications [1,2]. VN or V(C,N) precipitates contribute to ferrite grain refinement either by retarding the recrystallization of hot-deformed austenite or by acting as the heterogeneous nucleation sites for intragranular ferrite formation [1,3–10]. Although, the prime objective of V addition is to achieve precipitation strengthening from the fine VC or V(C,N) precipitates [1,3–6,11].

The nature, size and distribution of V-containing carbide, nitride or carbo-nitride precipitates depend not only on the V content but also on the C and N contents of the steel. Increase in both C and N levels in steel cannot be an effective way of strengthening as that can impart brittleness and hamper the weldability and formability. Therefore, either the steel can be richer in C or richer in N. Effect of C addition and N addition on the precipitation strengthening of V-microalloyed steels has been studied separately and both are found to be beneficial [1,3,7–9,12–29]. However, the effect of two different alloying strategies, i.e. high-C strategy and high-N strategy, on the microstructure and mechanical behaviour of V steels has hardly been compared.

V-microalloyed steels are being used both as flat-products like plates and sheets for linepipe and automotive applications and as long products like rods, bars and sections for construction applications [1,3,12–16,19–41]. Since the fine VC or V(C,N) precipitates primarily form during and after the austenite to ferrite transformation, hot-deformed V-microalloyed steels need to be cooled down at a very slow-rate over the transformation temperature range. Therefore, hot-rolled plates and sheets of V microalloyed steels are coiled over a temperature in the range of 500-750 °C [24,42–51]. The effect of coiling temperature on the precipitation in V containing steels has been studied independently for different steel compositions [24,42–52]. However, the effect of coiling temperature on the microstructure, precipitation and mechanical properties of steels having different alloying strategies (high-C or high-N strategy) has hardly been investigated and compared earlier. That sets the objective of the present study. Such a study not only fulfills the academic understanding on V precipitation in steel but also provides a guideline to the steel industries in selecting the C and N content and coiling temperature for V-microalloyed steel products.

2.      **Experimental details:**

Two laboratory cast blocks (20 Kg each) having different C and N contents were provided by Tata Steel Research & Development and Scientific Services Division, Jamshedpur. The blocks were cast following identical casting process. Chemical compositions of the steels are given in Table 1. High-C, low-N (HCLN) chemistry is typically used in the long-products like bars, rods and columns whilst, low-C, high-N (LCHN) chemistry is typically used in the flat products like plates, sheets and strips.

Cylindrical samples (10 mm diameter and 15 mm height) were prepared from the cast metal blocks and were subjected to thermo-mechanical processing simulation using Gleeble® 3800 simulator. The processing schedule applied in Gleeble is shown in Fig. 1. Samples were reheated to 1100 °C, soaked for 5 min, cooled down to 880 °C (cooling rate ~ 10 °C/s) and hot-compression tested at that temperature up to true strain, $\varepsilon = 1$ at a strain rate, $\acute{\varepsilon} = 1/s$. After compression testing, the samples were cooled down (cooling rate ~10 °C/s) to different temperatures in the range of 500–750 °C at 50 °C interval and isothermally held at that temperature for 1 h, followed by air-cooling. The isothermal holding treatment has been performed to simulate the industrial coiling treatment, where the steel cools down at a very slow rate from the coiling start temperature. Couple of thermocouples attached to the sample monitored the sample temperature during Gleeble® testing at an accuracy of ± 3 °C.

Cross-section of the compression tested samples near the thermocouple attached location was prepared following standard metallographic techniques. Microstructural study was conducted using optical microscope and scanning electron microscope (SEM) and image analysis was used for quantitative metallography. Grain area and equivalent circle diameter (ECD) grain size of 400-500 grains were measured from each sample to determine the average ferrite grain size. Thin foils were prepared by dimpling and ion-milling process and observed under JEOL-2100 model high resolution transmission electron microscope (HR-TEM) to study the fine V-precipitates. Solid cylindrical samples (4 mm outer diameter and 10 mm height) prepared from each cast block were austenitized at 1100 °C and continuously cooled down at 10 °C/s to the ambient temperature inside automated quench dilatometer (DIL 805A/D, TA Instruments), without applying any deformation.

Macro-hardness was measured at 20 Kgf load using a LV-700 model LECO® Vickers hardness tester as the average value of 5 readings from each sample. Micro-hardness indents were taken only from the ferrite regions of the investigated samples using UHL VHM-001 model micro-hardness tester (0.01 Kgf load) as the average value of twenty (20) readings from each sample. Rectangular samples (15 mm×20 mm×100 mm) were cut from the cast steel blocks and were subjected to a processing schedule similar to that shown in Fig. 1. A forging press was used for applying the compressive deformation and the isothermal holding treatment was carried out inside a muffle furnace. K–type thermo-couple was attached to the samples to monitor the temperature. Finally, tensile specimens were prepared from the forged blocks following ASTM E 8 standard [53] and tested at ambient temperature (25 °C) and cross-head velocity 0.5 mm /min using Instron® 8862 servo electric test system (10 t).

3. **Results and discussions:**

The investigated steels contain a moderate level of V addition (0.05 wt.%) as typically used in industrial HSLA steel grades for structural, linepipe and automotive applications. C and N levels are varied such that a comparative assessment can be made between two different alloy strategies, vanadium-carbon strategy and vanadium-nitrogen strategy. Thermo-Calc® software was used for the prediction of nature, fraction and stability of V-precipitates in the investigated steels, Fig. 2. As per the prediction, under equilibrium cooling condition, precipitation starts at higher temperature (~1050 °C) in low-carbon high nitrogen, LCHN, steel than that in high carbon low nitrogen, HCLN, steel (~1010 °C), Fig. 2. Therefore, at the soaking temperature of 1100 °C all the V precipitates are expected to be dissolved in austenite solution, Fig. 2. According to Thermo-Calc®, V-precipitates that form at higher temperatures in austenite (>850 °C) and at lower temperatures during austenite-ferrite transformation in HCLN steel are predicted to be richer in N and C, respectively. On the other side, primarily VN precipitates are expected to form in LCHN steel irrespective of precipitation temperature. This aspect is reported in detail in another study [54]. Precipitate fraction at the ambient temperature is predicted to be higher in HCLN steel than that in LCHN steel, Fig. 2.

3.1.   Transformation behaviour and microstructures of investigated samples:

Time-temperature-transformation (TTT) diagram of the investigated steels were predicted from JMatPro® software version 8.0, Sente software Ltd. The range of isothermal holding temperatures used in the present study are indicated by dotted lines on the TTT diagrams in Fig. 3(a,b). The cooling curves obtained from the dilatometric study showing the sample dilation during cooling from 1100 °C (at 10 °C/s cooling rate) without deformation are presented in Fig. 3c. Both the TTT diagrams and the cooling curves indicate that the ferrite transformation start temperature is ~50-70 °C higher and the incubation time required for the onset of ferrite transformation is ~ 8–10s lower in LCHN steel ($A_{r3}$~720 °C) than those in HCLN steel ($A_{r3}$~650 °C), Fig. 3. This is a clear indication that the decrease in C content reduced the hardenability and increased the transformation temperature of LCHN steel as compared to HCLN steel. Microstructures of the samples continuously cooled at 10 °C/s inside dilatometer are given in Fig. 4. Microstructure of HCLN steel showed the formation of polygonal ferrite along the prior-austenite grain boundaries and bainite and martensite at the prior-austenite grain interior locations, Fig. 4a. Polygonal ferrite formed at a higher quantity in LCHN steel sample, along with some acicular ferrite and upper bainite, Fig. 4b. The actual transformation behaviour of the investigated steels may not exactly follow Fig. 3 as the samples are hot-deformed (at 880 °C) before cooling and isothermal holding. Although Fig. 3 can provide a guideline to explain the differences in transformation behaviour and the resultant microstructural evolution in HCLN and LCHN steels as identical processing schedules were used for both the steels.

The optical micrographs of HCLN steel and LCHN steel samples are presented in Fig. 5 and Fig. 6, respectively, for different isothermal holding temperatures. Fractions of the microstructural constituents determined by the image analysis are presented in Table. 2. The microstructures of LCHN steel were dominated by ferrite, whilst, harder constituents such as, pearlite and bainite were present at significant proportion in HCLN steel, apart from the ferrite matrix, Fig. 5 and Fig. 6. The variation in average ferrite grain size and the total fraction of harder constituents (pearlite + bainite) with isothermal holding temperature is presented in Fig. 7.

Martensite was not present in the isothermally held samples. Average ferrite grain sizes were ~ higher in LCHN steel (7.7-8.9 µm) than that in HCLN steel (5.7-8.4 µm), Fig. 7a. This could be attributed to higher transformation temperature and lower incubation time for ferrite transformation in LCHN steel, which allowed greater opportunity for ferrite grain growth, as compared to HCLN steel. Total fraction of the harder constituents (pearlite + bainite) was ~ 20-30% higher in HCLN steel samples, due to its higher hardenability than LCHN steel, Fig. 7b. Ferrite fraction in HCLN steel remained within a close range, Table 2. Besides lamellar pearlite, degenerated pearlite was found in the HCLN samples isothermally held at and below 600 °C. Upper bainite was also found in HCLN steel samples that were held either at higher temperatures (700-750 °C) or at lower temperatures (500-550 °C), Table 2. On the other side, upper bainite formed only at higher holding temperatures (650 °C and higher) in LCHN steel.

The microstructures of the investigated samples can be explained from the transformation behaviour of the steels. Dilatometry study showed that under 10 °C/s cooling rate, allotriomorphic ferrite formation started in HCLN steel and LCHN steel at $Ar_3$ ~ 650 °C and $Ar_3$ ~ 720 °C, respectively, Fig. 3 and Fig. 4. Therefore, some amount of allotriomorphic ferrite formation was expected even during sample cooling when the isothermal holding temperatures were lower than the corresponding $Ar_3$ temperature. Now ferrite formation along the prior austenite grain boundaries during cooling can shift the carbon towards the austenite grain interior locations and influence the subsequent transformation at those locations during isothermal holding. This could have promoted the formation of degenerated pearlite and upper bainite at lower holding temperatures in HCLN steel, Table 2. The TTT diagrams in Fig. 3(a,b) indicate that at higher holding temperatures (700-750 °C) the decomposition of austenite possibly remained incomplete even after the prolonged isothermal holding and in such a scenario, pearlite and bainite transformation occurred from remaining austenite during the final cooling stage (after isothermal holding). This hypothesis could explain the formation of bainite at higher holding temperatures in both the steels, Table 2. Decomposition of austenite in both the investigated steels reached completion solely by the diffusional transformation at the intermediate holding temperatures (600–650 °C).

3.2. <u>Precipitate study in the processed samples:</u>

The bright field and dark field TEM images and selected area electron diffraction (SAED) analysis of the precipitates present in the ferrite matrix of HCLN steel and LCHN steel are given in Fig. 8 and Fig. 9, respectively. The bright field images of HCLN steel samples isothermally held at 500–650 °C show the presence of numerous fine precipitates, often interacting with the matrix dislocations, Fig. 8(a, b, d and e). SAED analysis indicated that the precipitates had FCC crystal structure with a lattice parameter of ~ 4.16 Å. The precipitates could possibly be the VC or V(C,N) precipitates as 4.16 Å is the lattice parameter of VC. In case of 600-650 °C holding temperatures, severe dislocation-precipitation interaction can be noticed,

Fig. 8(d-h). In general, precipitates were found to be distributed in a random fashion up to 650 °C, Fig. 8(a-h). At higher holding temperatures (700–750 °C), parallel arrays of interphase precipitation has been found and the precipitates were also coarsened into larger size, Fig. 8i. Dislocation density and dislocation-precipitation interaction in the ferrite matrix were comparatively lower at the higher holding temperatures (700–750 °C) than the intermediate holding temperatures (600–650 °C). The V(C,N) precipitates that formed upon holding at 700-750 °C had a lattice parameter of ~ 4.12 Å. Those V(C,N) precipitates were expected to be richer in N than C.

Interphase precipitation of VC or V(C,N) during austenite to ferrite transformation under slow-cooling condition or upon isothermal holding at high inter-critical temperature (~700 °C) is an established phenomena in V-microalloyed steels [30–33,55]. Lower transformation temperature increases the driving force for austenite to ferrite transformation and accelerates the interphase migration. As a result, interphase precipitation becomes difficult and random precipitation takes place. The geometrically necessary dislocations (GNDs) that form in ferrite during austenite to ferrite transformation provide the heterogeneous nucleation sites for the precipitation [1].

Compared to HCLN steel, the precipitates in LCHN steel were finer, more numerous and more uniformly distributed, Fig. 9. The dislocation density and the extent of precipitation-dislocation interaction in the ferrite matrix of LCHN steel was less severe than that of HCLN steel, Fig. 9(a-f). This can be attributed to the smaller fraction of harder microstructural constituents in LCHN steel, which reduced the density of GNDs in the adjacent ferrite matrix of LCHN steel as compared to that of HCLN steel. Fine interphase precipitation was found to occur at high holding temperature (750 °C) as indicated in Fig. 9g, while random distribution of precipitates was noticed at lower holding temperatures, Fig. 9(a-f). SAED analysis suggested that the precipitates could be richer in N than C, i.e.VN or V(C,N) precipitates, having a lattice parameter of ~ 4.12 Å.

Comparison of average precipitate size, number density and the volume fraction of the precipitates between HCLN and LCHN steels are plotted in Fig. 10. Decrease in isothermal holding temperature from 700 °C to 600 °C not only decreased the precipitate size and increased the precipitate density but also increased the precipitate volume fraction to a small extent, Fig. 10c. Further decrease in holding temperature to 500-550 °C refined the precipitate size but significantly reduced the precipitate volume fraction. The rate of increase in precipitate size was higher in HCLN steel than that of LCHN steel. These observations can be explained in view of the following thermodynamic and kinetic principals:

- The driving force for precipitation increases with the decrease in holding temperature, which is expected to increase the precipitate nucleation rate [56,57].
- At lower holding temperatures (500-550 °C) lack of thermal activation for V-diffusion could have affected the precipitation kinetics, which suppressed the precipitation process.

- High holding temperatures (700-750 °C) on the other side, provide greater thermal activation for the diffusion of V (slowest diffusing element), which increases the precipitate size by helping the precipitate growth and coarsening [56,57].
- Diffusional transformation of austenite to ferrite remained incomplete after isothermal holding at higher temperatures (≥ 700 °C) and it was followed by shear transformation during final cooling. Shear transformation also occurred during isothermal holding at lower temperatures (500–550 °C). At intermediate holding temperatures of 600–650 °C complete diffusional transformation of austenite was expected. Since V(C,N) precipitation is a diffusion controlled process and primarily occurs during diffusional transformation of austenite [1,29,33,36], the extent of precipitation was higher at intermediate holding temperatures.
- Better coarsening resistance of VN or N-rich V(C,N) precipitates as compared to VC or C-rich V(C, N) precipitates [3,19,36] led to finer precipitate size in LCHN steel than that in HCLN steel, especially at higher holding temperatures.

Similar to the present study, earlier studies also reported the formation of numerous fine microalloy precipitates (VC, NbC and even (Ti, Mo)$_2$C) at intermediate coiling temperatures (500–600 °C) not only in V-microalloyed steels but also in presence of other microalloying elements [24,42–51].

### 3.3. Hardness and tensile properties of the investigated samples:

Variation in Vickers macro-hardness (VHN) and micro-hardness (micro-hardness only from the ferrite matrix) for both the steel samples with isothermal holding temperature are presented in Fig. 11. Since macro-hardness and micro-hardness were determined under different loads, those readings cannot be compared directly as indentation size effect as well as the effect of local microstructure (i.e. ferrite grain size and presence of harder constituents) can influence the hardness values. One common trend however, can clearly be noticed from Fig. 11 that both macro-hardness and ferrite micro-hardness were higher (by ~20-24 VHN) in the samples isothermally hold at intermediate temperatures (600-650 °C) than the other holding temperatures, Fig. 11.

Tensile stress-strain curves of the samples hot-forged and isothermally held at 500 °C, 600 °C and 700 °C are shown in Fig. 12. Yield strength (YS) of HCLN and LCHN steel samples varied in the range of 520-650 MPa and 500-600 MPa, respectively, Fig. 12. Therefore, HCLN steel showed ~ 20-50 MPa higher YS than LCHN steel. LCHN steel on the other hand showed ~ 3-5 % higher ductility (total elongation) than that of HCLN steel, Fig. 12. Higher strength and lower ductility of HCLN steel can certainly be attributed to its higher C content and the resultant higher pearlite and bainite fraction in that steel (Fig. 7b), with respect to LCHN steel. The average ferrite grain size was also finer in HCLN steel, Fig. 7a. The strain hardening ability of

HCLN steel was much stronger than LCHN steel which resulted in significantly higher ultimate tensile strength (UTS) (by ~100–150 MPa) in HCLN steel than that of LCHN steel, Fig. 12.

3.4. Discussion on the effect of microstructure and precipitation on the tensile properties:

The contribution from different strengthening mechanisms on the overall yield strength of ferritic steels can be given by the following equation:

$$\sigma_y = \sigma_0 + \sigma_{ss} + \sigma_{gb} + \sigma_d + \sigma_p \qquad (1)$$

where, $\sigma_y$ is the yield strength of the steel, $\sigma_0$ is the lattice friction stress (~ 48 MPa), $\sigma_{ss}$, $\sigma_{gb}$, $\sigma_d$ and $\sigma_p$ represent solid solution strengthening, grain boundary strengthening, dislocation strengthening and precipitation strengthening, respectively. Strengthening contribution of each mechanism can be determined from a set of equations that are extensively reported in earlier studies [58].

Chemical composition in the ferrite matrix corresponding to the isothermal holding temperature of the sample has been predicted from the Thermo-Calc® software and was used to estimate the solid solution strengthening, $\sigma_{ss}$. Investigated steels showed similar contributions from $\sigma_{ss}$ (70-80 MPa). Grain boundary strengthening, as determined from the average ferrite grain sizes (given in Fig. 7a) showed ~ 30-40 MPa higher contribution in HCLN steel than that in LCHN steel. Higher dislocation density (estimated from the TEM images following established procedure [59]) further adds to higher dislocation strengthening (by ~20-30 MPa) in HCLN steel than that in LCHN steel. Besides the above mentioned effects, a significant strengthening contribution is expected to arise from the higher fraction of harder microstructural constituents (pearlite and bainite) in HCLN steel than that in LCHN steel. In spite of the greater strengthening contributions coming from the different sources in HCLN steel, its YS was not significantly higher than the YS of LCHN steel, Fig. 12. Study on the precipitation strengthening can help in explaining this difference.

Strengthening contribution of incoherent precipitates by dislocation looping mechanism can be determined from the well known Ashby- Orowan equation [60] as stated below:

$$\sigma_p = \frac{5.9\sqrt{f}}{\bar{x}} \ln \left[\frac{\bar{x}}{2.5 \times 10^{-4}}\right] \qquad (2)$$

where, $\bar{x}$ is the average precipitate size (in µm) and $f$ is the volume fraction of precipitates. The values of $\bar{x}$ and f can be obtained from Fig. 10. The calculated $\sigma_p$ and $\sigma_y$ values of the simulated samples for different holding temperatures are plotted in Fig. 13. According to Fig. 13 higher precipitation strengthening (by ~30-50 MPa) is expected in LCHN steel which can be attributed to its finer precipitate size and higher precipitate density, compared to HCLN steel. The overall strength of ferrite is however, predicted to be higher in HCLN steel due to the greater contributions from grain boundary strengthening and dislocation strengthening, with respect to LCHN steel. Fig. 13 also justifies the higher hardness and strength in the samples held at intermediate temperatures (600–650 °C), which can be attributed to the higher precipitation strengthening contribution, as compared to the other holding temperatures. Higher strain hardening ability in HCLN steel can be attributed to the stronger dislocation-precipitation interaction in that steel (mentioned in Section 3.2). The trend shown by the strengthening calculations are therefore, in line with the experimental results except for the fact that the exact strength levels and the difference in strength of the investigated steels could not be predicted accurately. This can be attributed to the following factors:

- It is difficult to estimate the exact strengthening contributions of the harder constituents like pearlite and bainite, in presence of ferrite.
- Presence of harder microstructural constituents can result in preferential strain partitioning towards the softer ferrite matrix, which controls the yielding of softer ferrite matrix [61,62].
- Precipitation strengthening contribution can be different for random precipitation and interphase precipitation [30].
- Ashby-Orowan equation (eqn. 2) for the prediction of precipitation strengthening is valid only in case of incoherent precipitates, but it is invalid for coherent and shearable precipitates [51], which are very fine (say, < 3 nm [56]) and are difficult to detect and quantify.

4. Conclusions:

In order to understand the effect of C and N levels and isothermal holding temperature on the microstructure, precipitation and mechanical properties of V-containing steels (0.05 wt.%), two different V-steels having different C and N levels, namely high-C low-N (HCLN) and low-C high-N (LCHN), were subjected to isothermal holding treatment over a temperature range of 500–750 °C after hot-deformation and finally air-cooled. Major conclusions derived from the study are listed below:

- Maximum precipitation strengthening has been found at intermediate isothermal holding temperatures (600 – 650 °C) in both the investigated steels as a result of the formation of fine V(C,N) precipitates at high density.
- In spite of the significantly smaller fraction of pearlite and bainite (by ~ 20-30 %), coarser average ferrite grain size (by ~ 2.0 – 2.5 µm) and lower dislocation density in the ferrite matrix, the yield strength of LCHN steel was close to HCLN steel. This can be attributed to the higher precipitation strengthening (by ~ 30–50 MPa) resulted from finer V-precipitates in LCHN steel than that of HCLN steel.

- Average V (C, N) precipitate sizes were finer in LCHN steel than that of HCLN steel, irrespective of the isothermal holding temperature. This can be due to the better coarsening resistance of VN, or N-rich V(C, N), precipitates (present in LCHN steel) as compared to VC, or C-rich V(C, N), precipitates (present in HCLN steel).
- Presence of harder constituents and the stronger interaction between incoherent V(C, N) precipitates and the geometrically necessary dislocations contributed to greater strain-hardening ability in HCLN steel with respect to LCHN steel.

Present findings indicate that low-C high-N composition in V steel is suitable for flat products (thinner plates and sheets) that are subjected to industrial coiling treatment and require good combination of strength and ductility. The beneficial effect of intermediate coiling temperature (600 – 650 °C) is evident. In a separate study, high-C low-N composition are found to be suitable for long products (rod and bars) and thicker plates, which are continuously cooled down after rolling and primary required to achieve the specified strength level [54].

**Acknowledgements:**

The authors duly acknowledge the financial support received from the Research & Development and Scientific Services Division of Tata Steel, Jamshedpur, experimental facilities offered by the Department of Metallurgical and Materials Engineering, Central Research Facility and Steel Technology Centre at IIT Kharagpur and equipment grant provided by Sponsored Research and Industrial Consultancy, IIT Kharagpur through SGIRG scheme. The authors would also like to sincerely thank Mr. S. Neogi and Dr. G.K. Dey from the Materials Science Division of Bhabha Atomic Research Centre, Mumbai for their help and support in performing some TEM studies for this work.

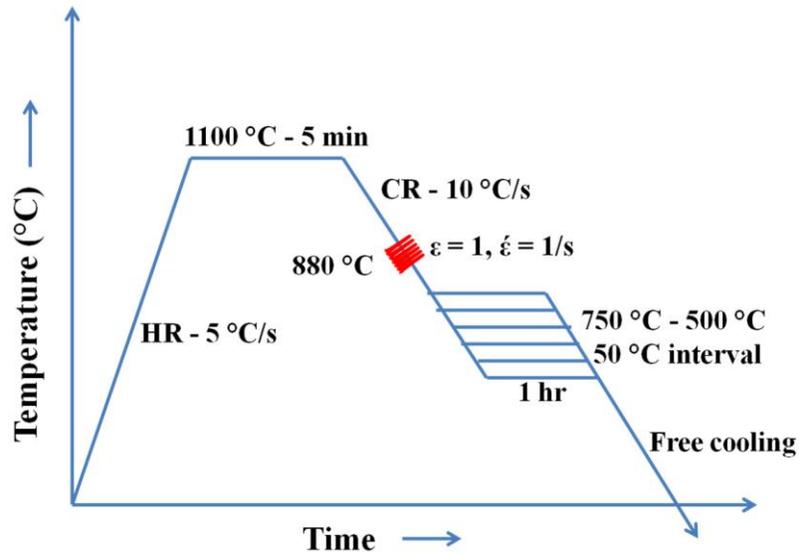

Fig. 1: Schematic diagram of the thermo-mechanical processing schedule applied in Gleeble® simulator. Abbreviations: HR: heating rate; CR: cooling rate.

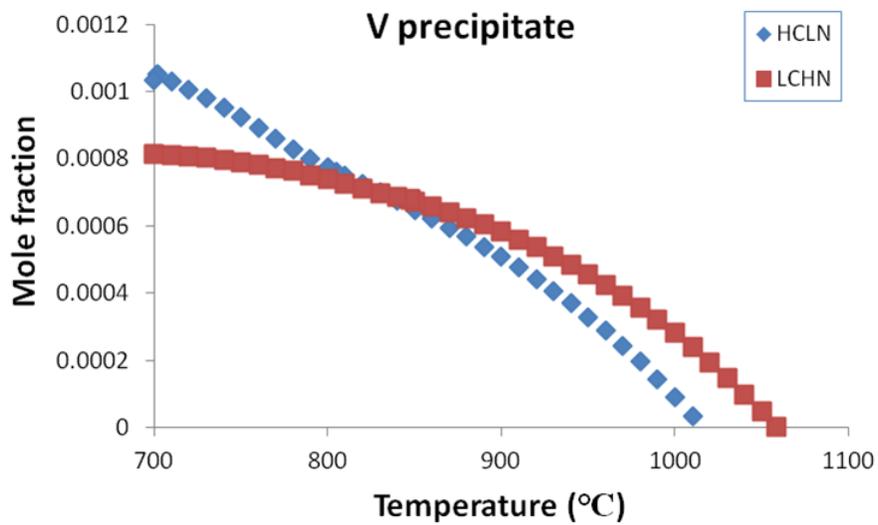

Fig. 2: Thermo-Calc® software prediction of equilibrium precipitate stability in the investigated steels.

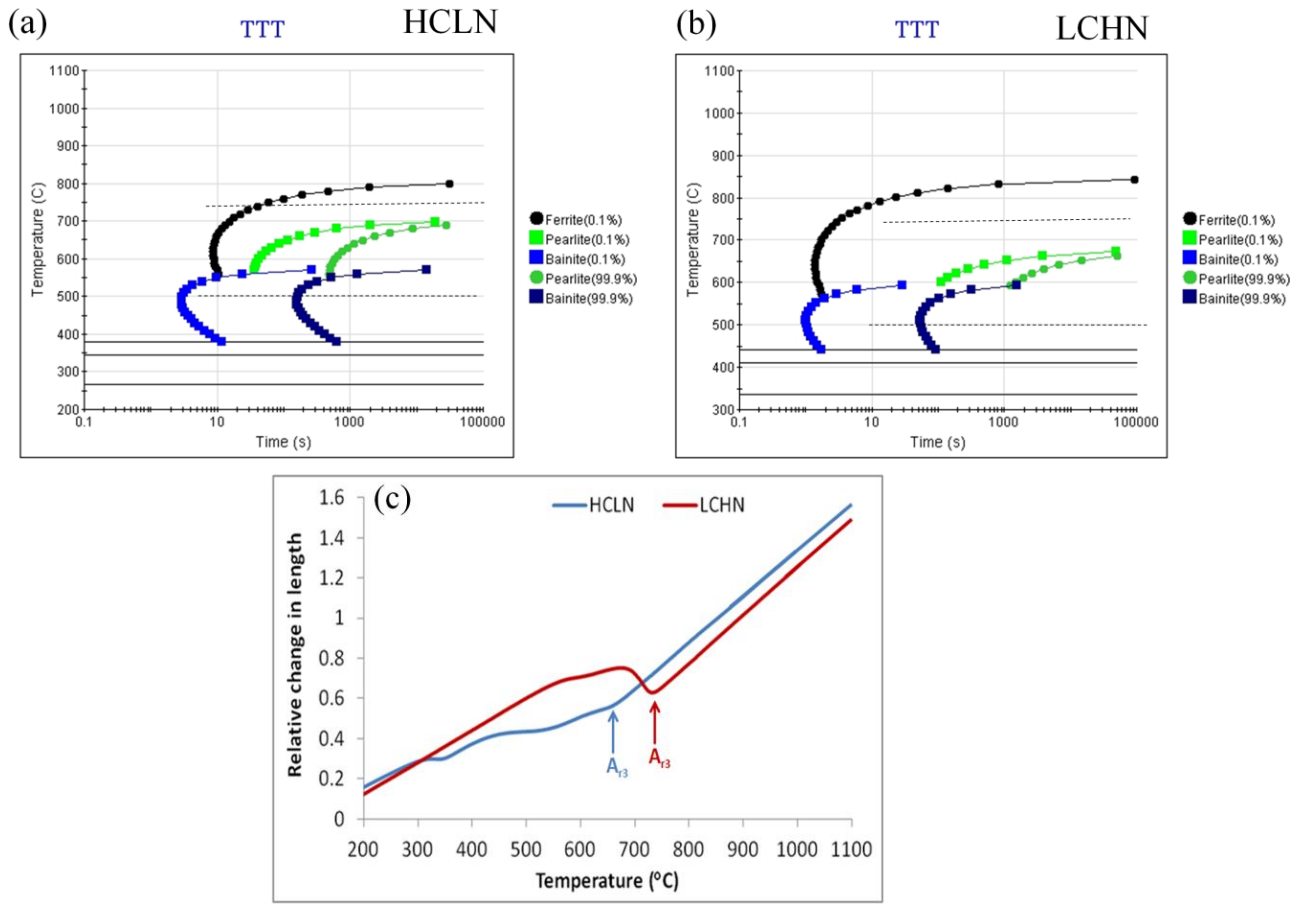

Fig. 3: TTT (Time temperature transformation) diagrams of (a) HCLN and (b) LCHN steels as predicted from JMatPro® software. (c) Dilation vs. temperature curves of the investigated samples upon cooling at a rate of 10°C/s. Austenite to ferrite transformation start temperatures during cooling ($Ar_3$) are indicated by arrows.

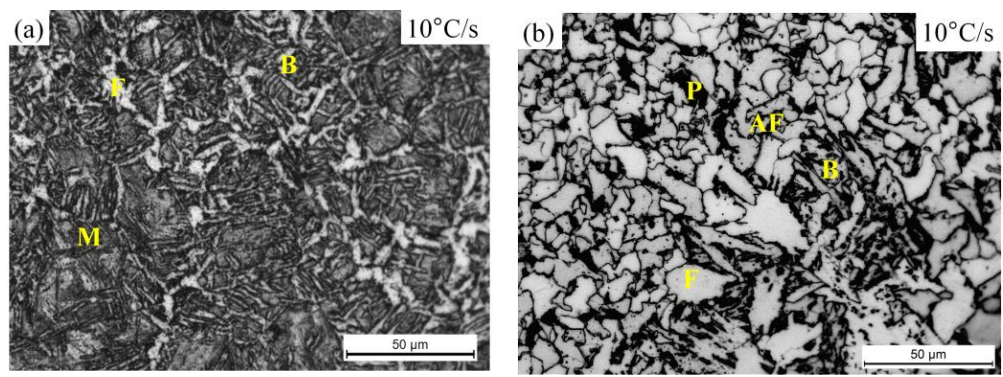

Fig. 4: Optical micrographs of dilatometry tested samples of (a) HCLN and (b) LCHN steels for the cooling rate of 10°C/s. Abbreviations: F: Polygonal ferrite, AF: Acicular ferrite, P: Pearlite, B: Bainite and M: Martensite.

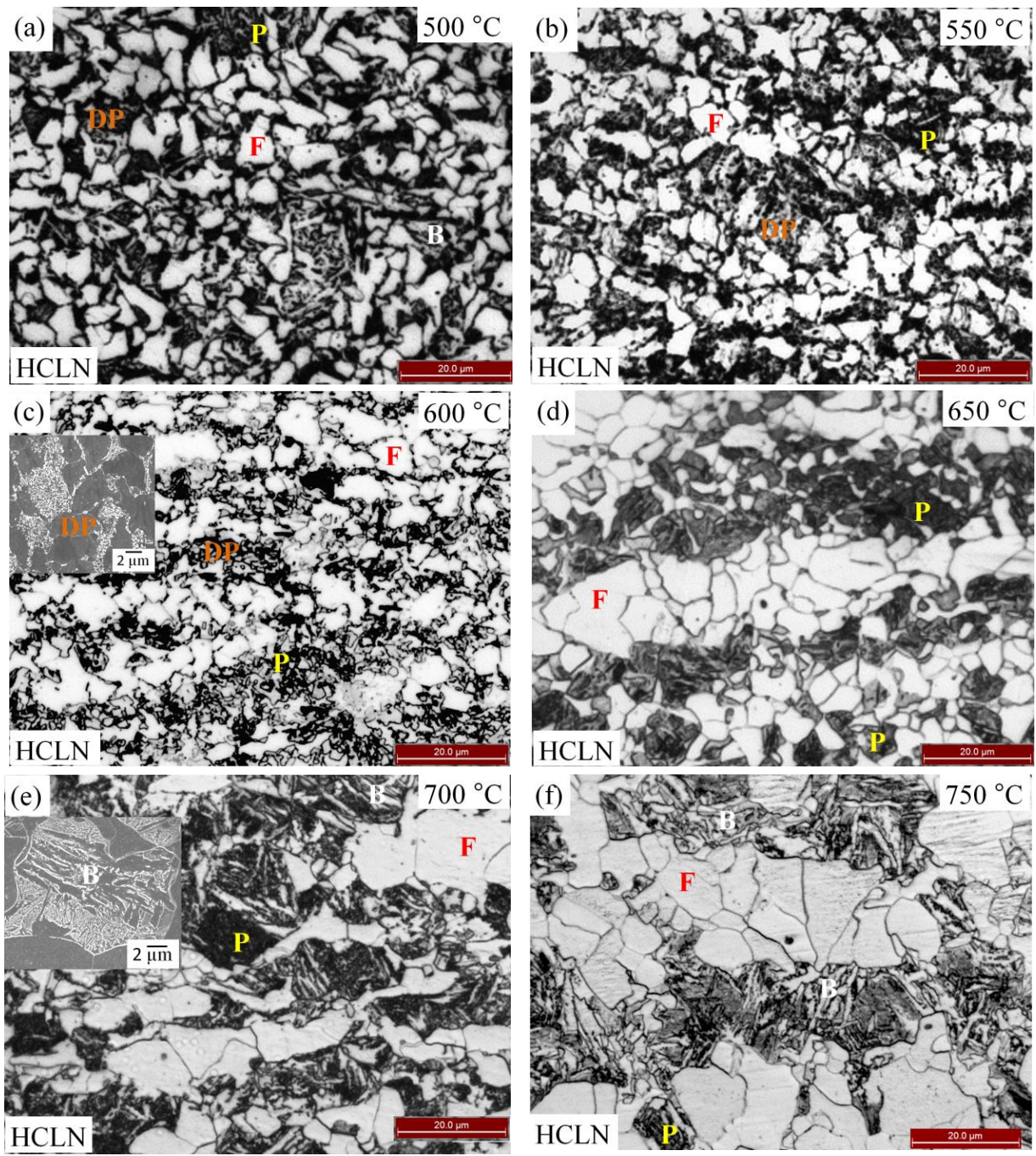

Fig. 5: Optical micrographs of HCLN steel samples for different isothermal holding temperatures as mentioned on the images. Abbreviations: F: Polygonal ferrite; P: Pearlite, B: Bainite and DP: Degenerated pearlite. High magnification scanning electron micrographs of degenerated pearlite and bainite are shown in the inset of (c) and (e) respectively.

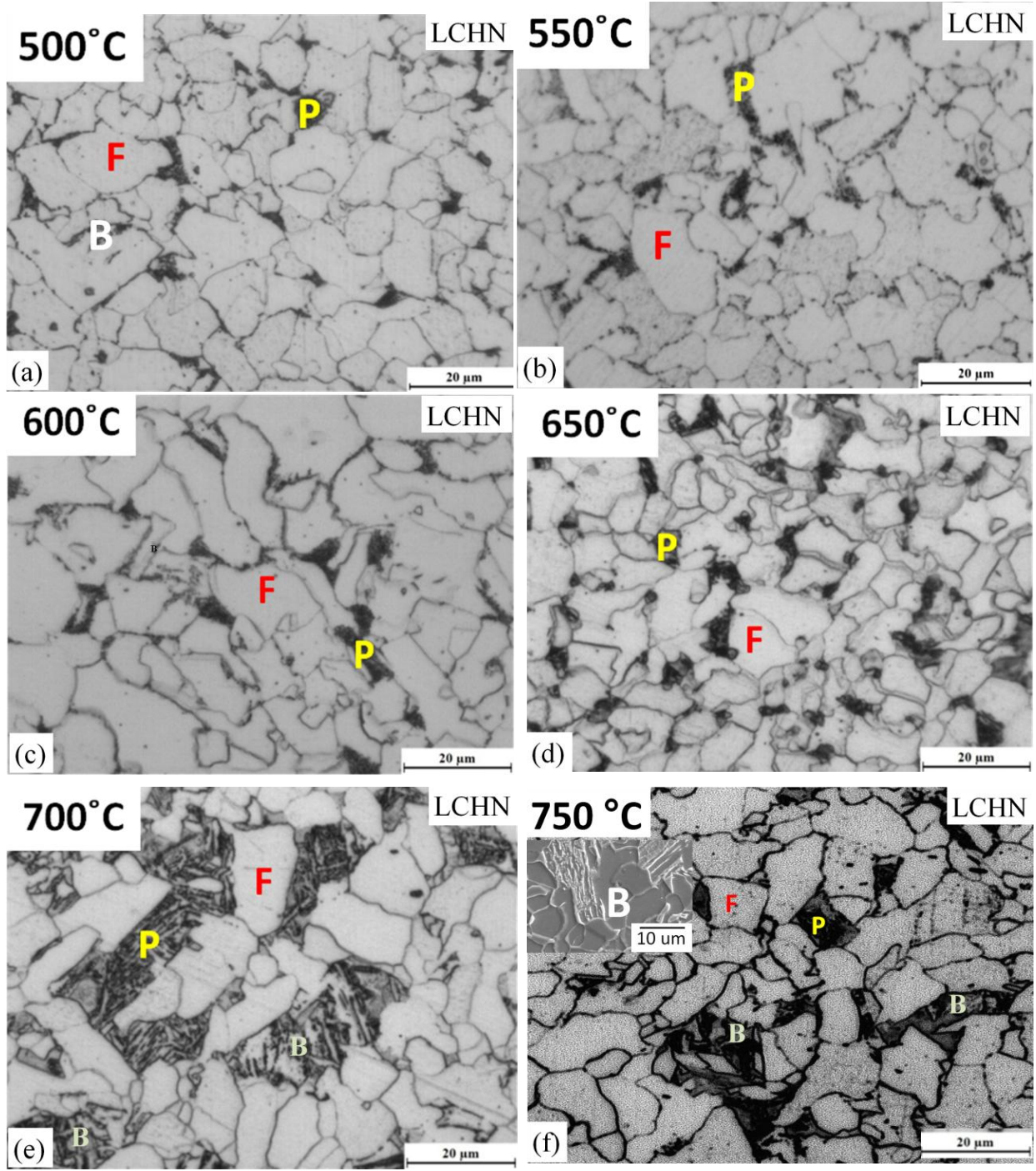

Fig. 6: Optical micrographs of LCHN steel samples for different isothermal holding temperatures as mentioned on the images. Abbreviations: F: Polygonal ferrite; P: Pearlite, B: Bainite and DP: Degenerated pearlite. High magnification scanning electron micrographs of degenerated pearlite is shown in the inset of (f) respectively.

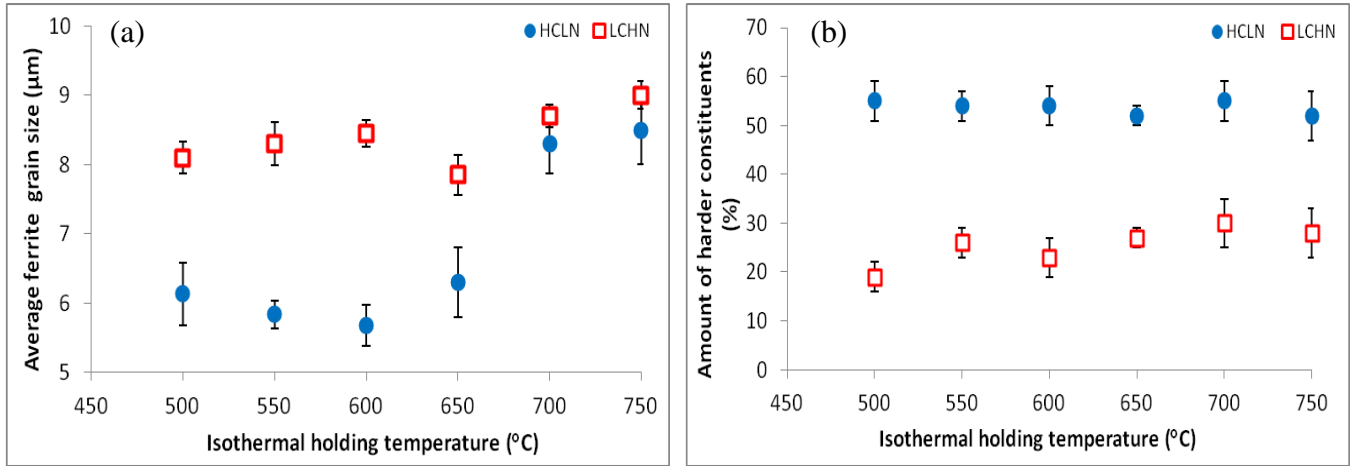

Fig. 7: Variation in (a) the average ferrite grain sizes and (b) the amount of harder constituents (pearlite and bainite) as the function of isothermal holding temperatures for the investigated steels.

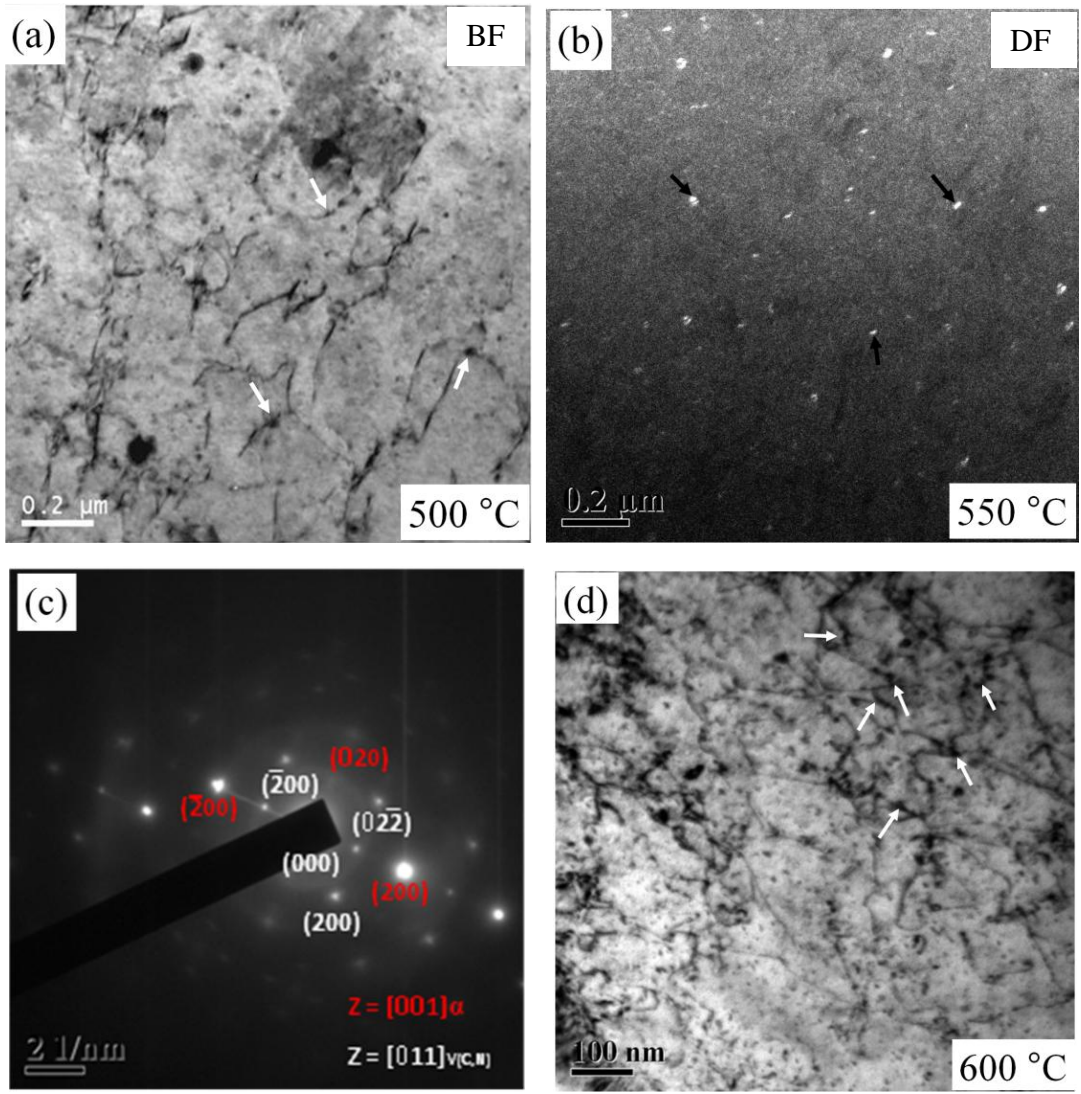

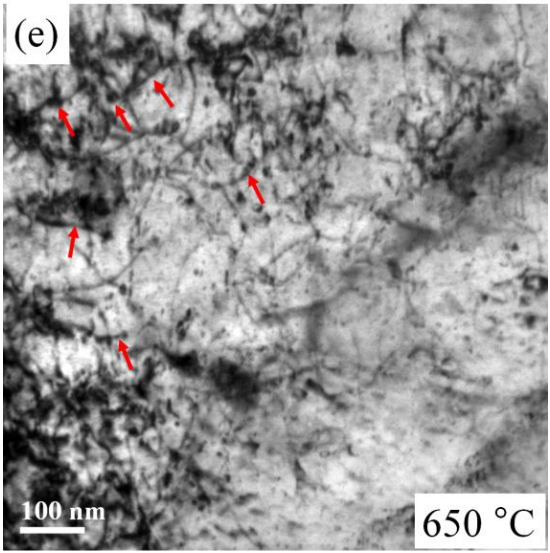
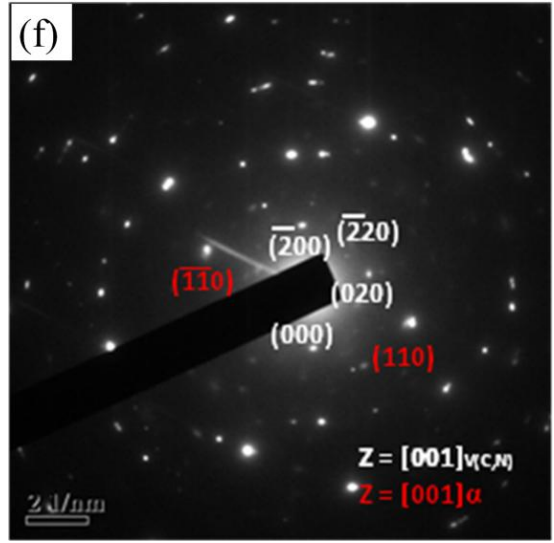
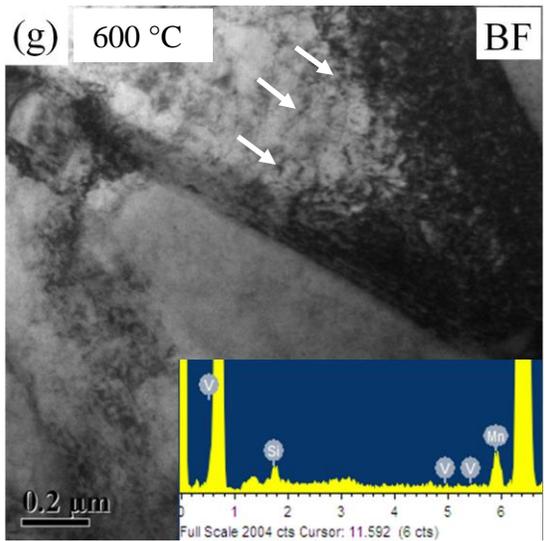
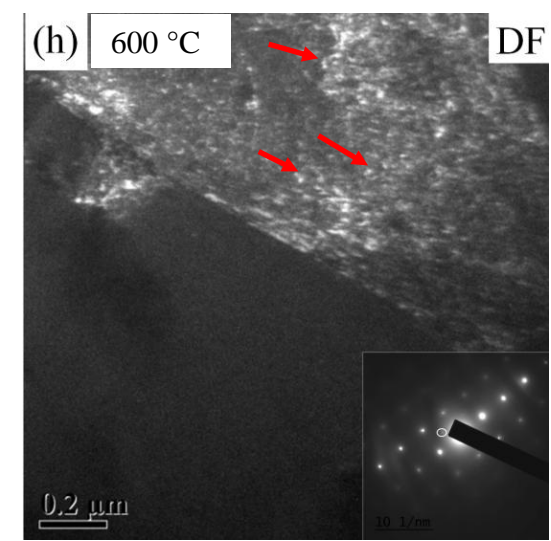
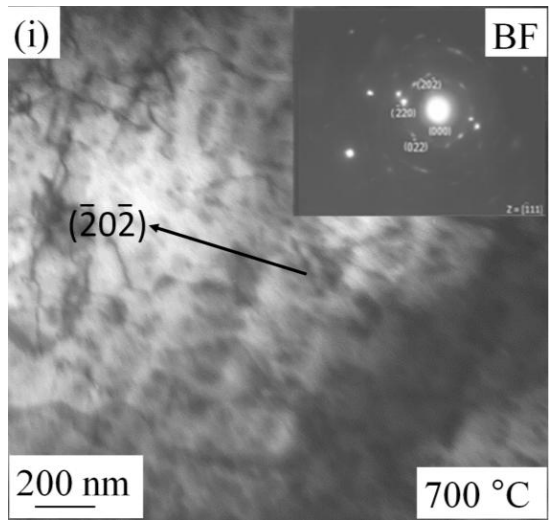

Fig. 8: Bright field (BF) and dark field (DF) transmission electron micrographs (TEM) and the corresponding SAED pattern analysis of the precipitates in the ferrite matrix of HCLN steel for different isothermal holding temperatures as mentioned on the micrographs. Energy dispersive spectroscopy analysis (EDS) of the fine precipitates is inserted in (g). Dislocations, precipitates and their mutual interactions are indicated on the micrographs by arrows. Interphase precipitates are shown in (i) and the interphase growth front is marked by an arrow.

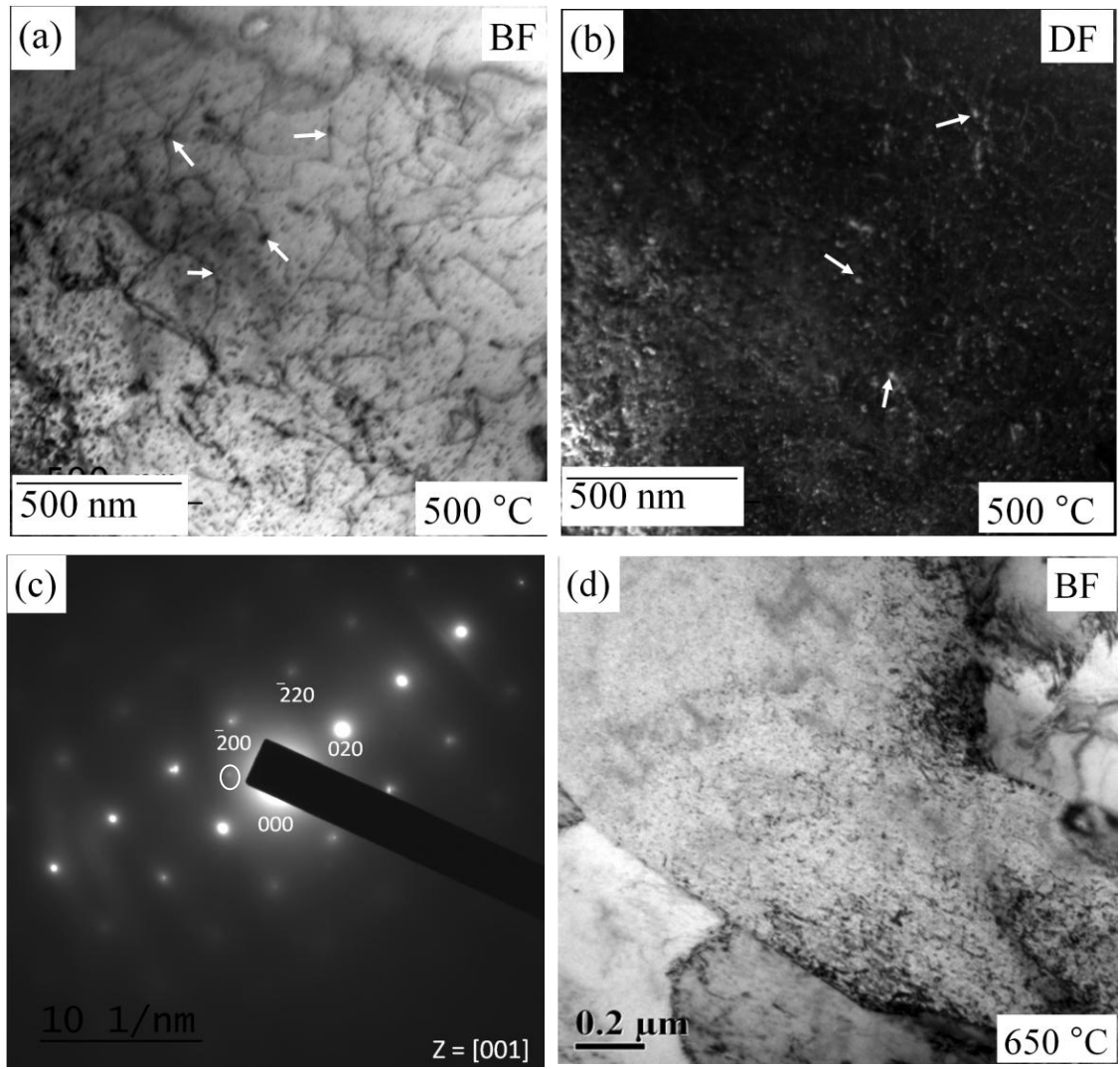

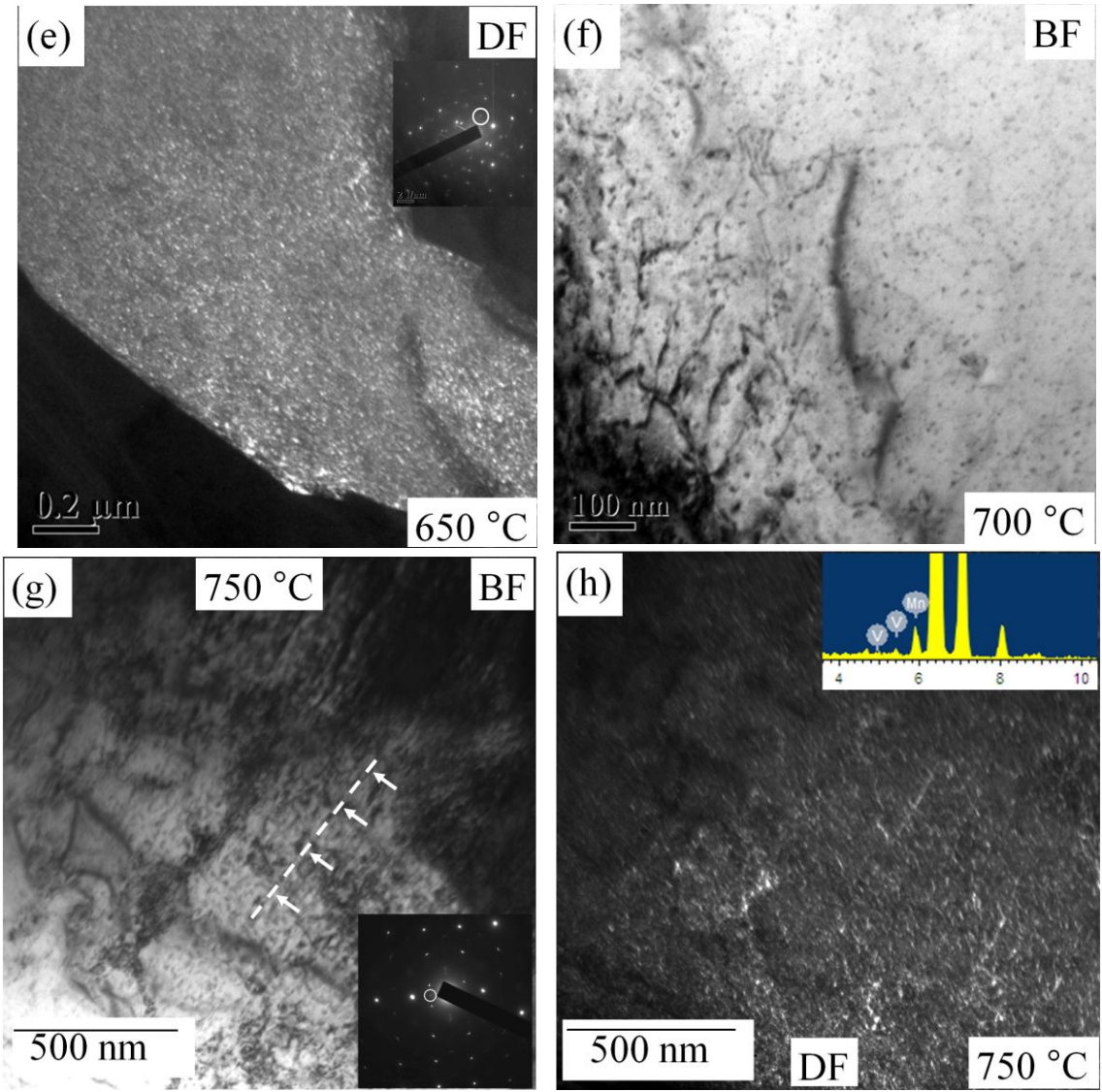

Fig. 9: Bright field (BF) and dark field (DF) transmission electron micrographs (TEM) and the corresponding SAED pattern analysis of the precipitates in the ferrite matrix of LCHN steel for different isothermal holding temperatures as mentioned on the micrographs. Energy dispersive spectroscopy analysis (EDS) of the fine precipitates is inserted in (h). Dislocations, precipitates and their mutual interactions are indicated on the micrographs by arrows. Interphase precipitates are shown in (g), where one of the precipitate arrays is indicated.

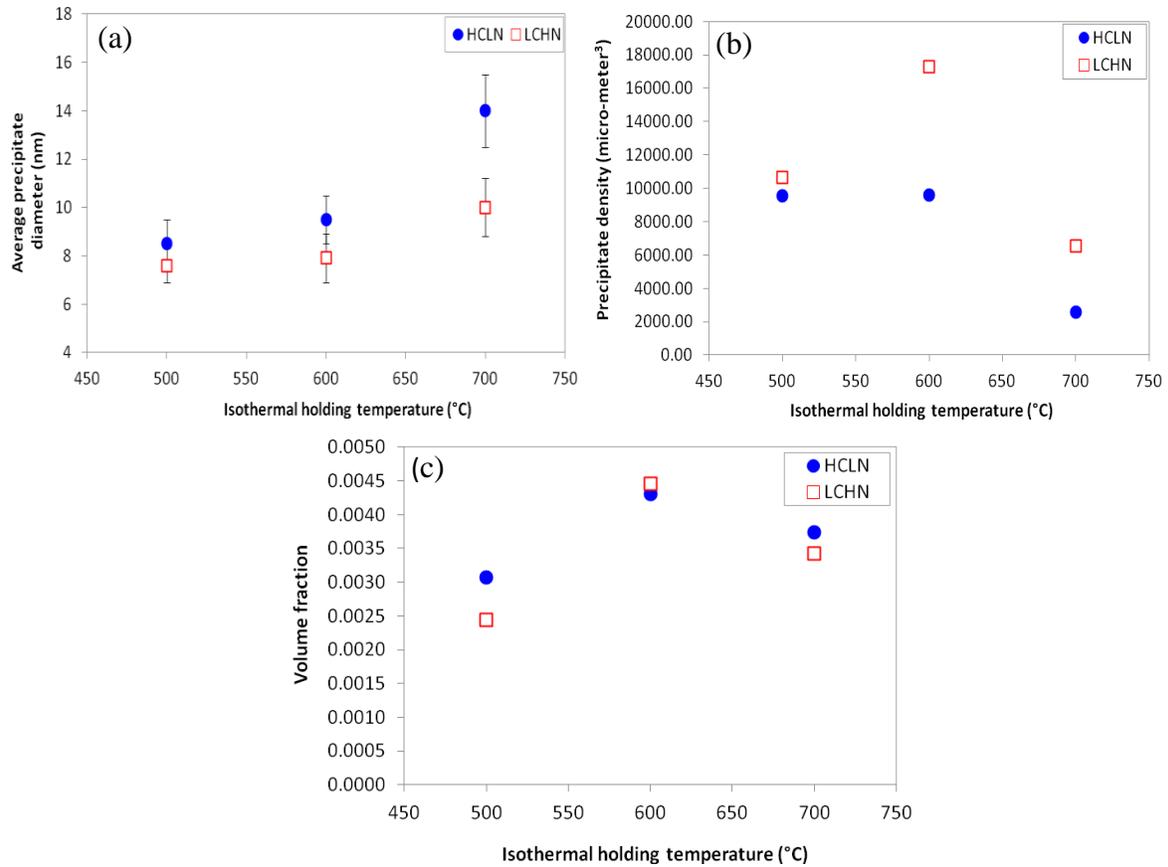

Fig. 10: Variation in (a) average precipitate size, (b) precipitate density and the (c) precipitate volume fraction with isothermal holding temperature for the investigated steels.

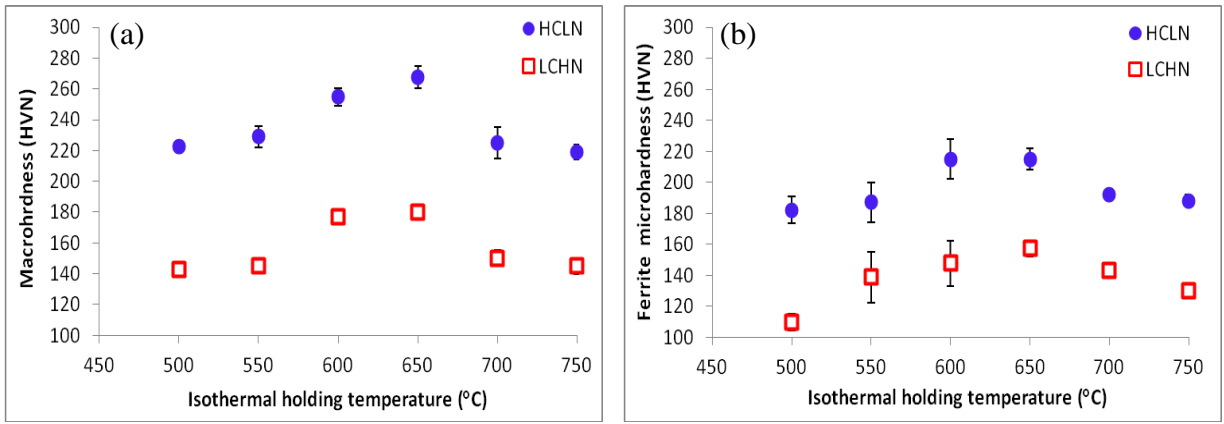

Fig. 11: Variation in (a) average macro-hardness and (b) average micro-hardness (from ferrite regions) of the investigated steels as the function of isothermal holding temperature.

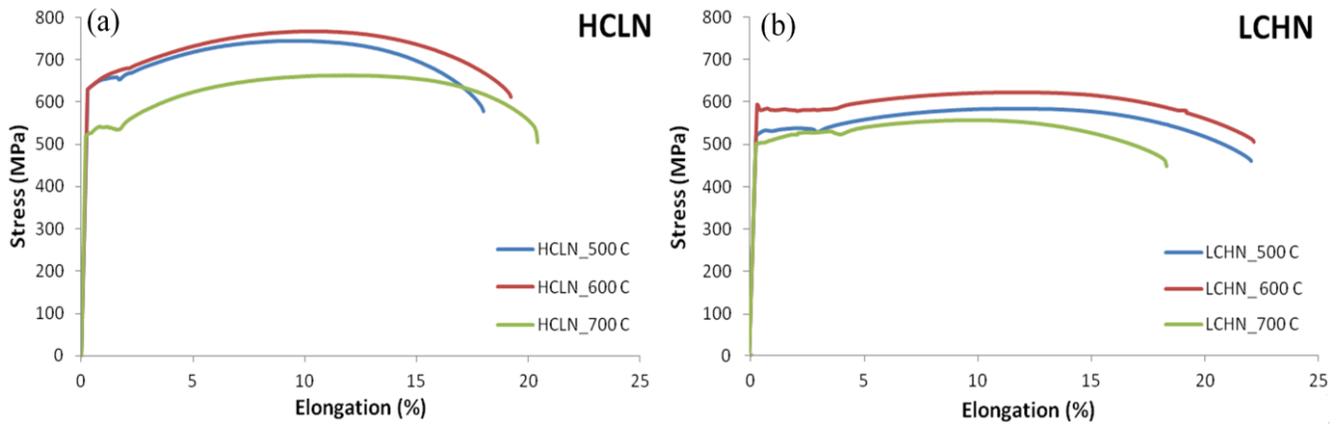

Fig. 12: Tensile stress-strain curves of (a) HCLN and (b) LCHN steel samples isothermally held at 500°C, 600°C and 700°C.

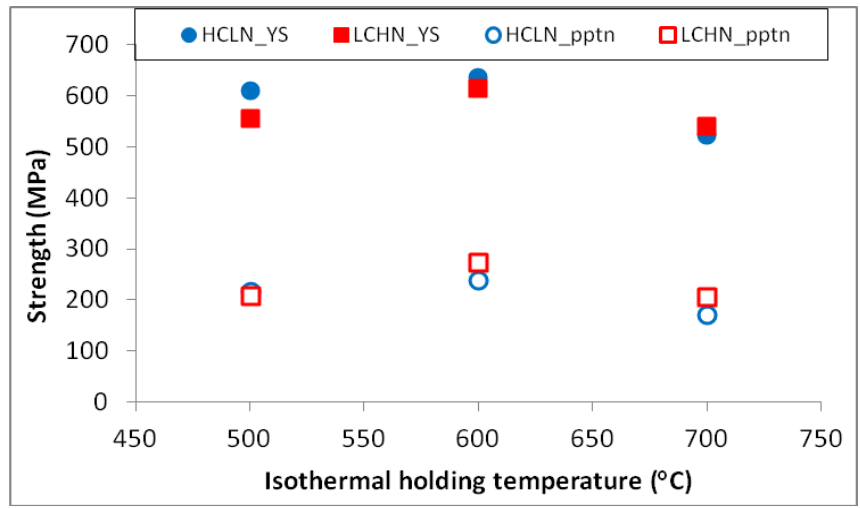

Fig. 13: Estimated yield strength levels (HCLN_YS and LCHN_YS) and precipitation strengthening contributions (HCLN_pptn and LCHN_pptn) of the investigated steels at different isothermal holding temperatures.

Table.1: Chemical composition (wt.%) of the investigated steels.

|      | C     | Mn   | Si   | S     | P    | V    | N      |
|------|-------|------|------|-------|------|------|--------|
| HCLN | 0.22  | 1.7  | 0.43 | 0.02  | 0.02 | 0.05 | 0.008  |
| LCHN | 0.055 | 1.65 | 0.42 | 0.009 | 0.01 | 0.05 | 0.0137 |

Table 2: Fractions (percentage) of different microstructural constituents in the thermo-mechanically simulated samples for different isothermal holding temperatures.

| Isothermal holding temperatures (°C) | HCLN | | | | LCHN | | |
|---|---|---|---|---|---|---|---|
| | Ferrite (%) | Pearlite (%) | Degenerate pearlite (%) | Bainite (%) | Ferrite (%) | Pearlite (%) | Bainite (%) |
| 500 | 45±3 | 25±3 | 10±3 | 20±3 | 81±3 | 14±3 | 5±2 |
| 550 | 46±2 | 29±1 | 25±2 | - | 79±4 | 26±4 | - |
| 600 | 46±1 | 30±3 | 24±1 | - | 77±2 | 20±3 | - |
| 650 | 48±3 | 52±3 | - | - | 73±3 | 27±3 | - |
| 700 | 45±1 | 25±4 | - | 30±3 | 70±1 | 20±3 | 10±1 |
| 750 | 48±3 | 17±1 | - | 35±3 | 72±2 | 12±4 | 16±2 |